\documentclass[fleqn,10pt]{arXiv}
\pdfoutput=1

\usepackage[english]{babel}
\usepackage{natbib}
\usepackage{float}
\usepackage{textcomp}

\definecolor{color1}{RGB}{120,0,0}
\definecolor{color2}{RGB}{0,20,20}
\definecolor{color3}{RGB}{0,0,0}

\usepackage{hyperref}
\hypersetup{hidelinks,colorlinks,breaklinks=true,urlcolor=color2,citecolor=color1,
            linkcolor=color1,bookmarksopen=false,pdftitle={Title},pdfauthor={Author}}

\JournalInfo{Reply}
\Archive{}

\PaperTitle{Not even wrong: Reply to Loreau and Hector \\
\vspace{0.25cm} \normalsize{Reply to ``Not even wrong: Comment'' by Loreau and Hector (2019)}\thanks{This preprint represents our official reply to the \citet{loreau.hector_19} Comment. A version of this reply was rejected for publication in \textit{Ecology} for non-scientific reasons.}}
\PaperSubtitle{Reply to Loreau and Hector Comment}

\Authors{Pradeep Pillai\textsuperscript{1,$\dagger$} and Tarik C. Gouhier\textsuperscript{1}}
\affiliation{\textsuperscript{1}\textit{Marine Science Center, Northeastern University, 430 Nahant Road, Nahant, MA 01908}}
\affiliation{\textsuperscript{$\dagger$}\textbf{Corresponding author}: pradeep.research@gmail.com}
\newcommand{\keywordname}{Keywords}

\Abstract{The \citet{loreau.hector_19} Comment on our paper \citep{pillai.gouhier_19} failed to address the two core elements of our critique, both the circularity of the BEF research program, in general, and the mathematical flaws of the Loreau-Hector partitioning scheme, in particular. Loreau and Hector avoided dealing with the first part of our critique by arguing against a non-existent claim that all biodiversity effects could be reduced to coexistence, while the mathematical flaws in the Loreau-Hector partitioning method that we described in the second part of our critique were ignored altogether. Here, we address these misconceptions and demonstrate that all of the claims that were made in our original paper hold. We conclude that (i) BEF studies need to adopt baselines that account for coexistence in order to avoid overestimating the effects of biodiversity and (ii) the LH partitioning method should not be used unless the linearity of the abundance-ecosystem functioning relationship in \emph{monocultures} can be verified for all species.}
\begin{document}
\flushbottom
\maketitle
\thispagestyle{firstpage}

%----------------------------------------------------------------------------------------
%	ARTICLE CONTENTS
%----------------------------------------------------------------------------------------

\section*{Introduction}

We awaited the \citet{loreau.hector_19} Comment on our recent paper \citep{pillai.gouhier_19} in the hopes that it would offer the opportunity for a back-and-forth that could unpack some of the hidden assumptions and beliefs that remained unexamined in the biodiversity-ecosystem (BEF) research program over the last three decades. Unfortunately, neither of the two components of our critique -- the circular nature of the BEF research program, nor the mathematical flaws of the Loreau-Hector (LH) partitioning method -- were addressed at all in Loreau and Hector's Comment. Instead of responding to our first point, Loreau and Hector argued against a non-existent claim that all BEF effects could be reduced to coexistence. Our second point regarding the flawed mathematics of the LH partitioning method was simply dismissed out-of-hand with a mixture of evasion and subterfuge. In place of a substantive response to our critique, much of the Loreau and Hector Comment was simply given over to boosterism: an exercise in declaiming both the historical successes of the BEF research program, in general, and Loreau and Hector's ``classic'' partitioning method, in particular.

As a result, we are forced to provide a conceptual summary of the main points of our original critique, while addressing, along the way, Loreau and Hector's misreading of our paper, and consequently, their misdirected arguments \citep{loreau.hector_19}. First, it is important to reiterate that our paper contained two distinct critiques of the BEF research program that need to be distinguished from each other. Although each stands on its own, both claims have been confounded to some degree in the initial response to our paper. The two claims are as follows:

\textbf{Claim 1}. There is an underlying circularity to the BEF approach -- whether we are talking about absolute ecosystem functioning, or net biodiversity effects on functioning. Biodiversity in the form of species coexistence or co-occurrence already in itself logically entails, \emph{to some degree}, a positive BEF relationship -- it is built into the assumptions of how ecological communities exist \emph{ceteris paribus}, and is thus, (again, \emph{in part}) a confounding artefact of the experimental design (there can be no BEF experiments without species coexisting).

\textbf{Claim 2}. The dominant BEF statistical approach based on the LH method (and all its extensions over the last two decades) are mathematically flawed. The mathematical issues we elucidated render the two key measures of the LH approach ineffectual for inferential purposes: (i) the net biodiversity effect itself and (ii) its partitioning into selection and complementarity.

Neither of the above two claims requires the other to hold, but they do reinforce each other. LH completely misconstrued the point of the first claim and largely ignored the second claim. Below we will attempt to address some of the misconceptions.

\section*{Claim 1: The circularity of BEF measures}
LH appear to interpret our claim that there is a ``trivial element'' or ``artefactual'' component to current BEF measures as meaning that we attribute all observed BEF effects to coexistence, or that there is no ecosystem effect beyond mere coexistence. These incorrect interpretations were reinforced by the use of truncated quotes that stripped our statements of their context and altered their intended meaning (e.g., insinuating that we claimed that all measured biodiversity effects are an ``artefact of coexistence'' by stripping away clauses like ``at least in part'', ``some degree of'', or ``[a] trivial element''). The point made throughout our paper was that BEF measures combine a trivial element associated with species merely co-occurring together, since they are, at least in part, measuring an \emph{existential feature of communities or ensembles}. Thus, without accounting for the confounding effects of co-occurrence (which is required to conduct biodiversity experiments in the first place), BEF studies will have a built-in circularity.

LH claim that the circularity described above was far from obvious or trivial when the BEF research program was in its infancy because ``[n]iche partitioning and stable coexistence in species-rich communities were so little established at that time that the first wave of debate that stirred the BEF research field during the 1990s and early 2000s was precisely on the issue whether [\emph{sic}] biodiversity effects on ecosystem functioning could be entirely due to selection effects, that is, due to which species came to dominate mixtures in the absence of any niche differentiation (Huston 1997)''. This statement is misleading because it makes it seem as though niche-based coexistence was a relatively unknown concept in community ecology in the 1990s rather than the default expectation. The truth of the matter is that seminal experimental \citep{gause_34} and theoretical \citep{volterra_28,lotka_32} studies cemented the role of niche-based coexistence as one of the central organizing principles of community ecology early in the $20^\text{th}$ century, decades before the BEF research program was born. Alternative explanations for coexistence such as the intermediate disturbance hypothesis \citep{huston_79} and neutral theory \citep{hubbell_01} were controversial precisely because they challenged the dogma of niche-based coexistence.

Although the role of diversity in promoting ecosystem functioning was vigorously contested in the early days of the BEF research program \citep{huston_97, huston_ea_00}, the debate focused on whether the increase in ecosystem functioning observed in species-rich mixtures was due to diversity \emph{per se} or the sampling effect, which is associated with the increased probability of selecting productive species in more species-rich mixtures. The debate was not fundamentally about the drivers of coexistence or the relationship between niche-based coexistence and ecosystem functioning. Hence, when we stated that ``it is now widely recognized that niche partitioning is common in nature'' and that ``the presence of a positive relationship in most BEF studies is thus unsurprising and largely trivial'' because ``the functioning measured in BEF experiments likely represents, at least in part, a sort of redundant measure or artefact of coexistence itself'', we were not ``ignor[ing] the history of [our] science'' as LH incorrectly claimed but embracing one of its core concepts and organizing principles. That is why we designed our baseline to account for (rather than ignore) the effects of coexistence on ecosystem functioning in pairwise mixtures.

Overall, LH's misdirected line of attack on our argument about triviality occurs in a two-step process. First, LH misread our attempt to control for the `trivial' effects that coexistence has on ecosystem properties as us claiming that \emph{all} the effects of coexistence are trivial. Second, since we were purportedly claiming that all effects of coexistence are trivial, LH took us to task for only deleting the effects of pairwise coexistence and not all higher order effects, although doing so would result, as they pointed out, in the absurd case of all ecosystem effects being deleted. Clearly we never argued for anything quite that bizarre. We now identify the sources of these serious misunderstandings.

\subsection*{Misapprehending the role of pairwise effects}
One of the reasons why LH believed that we wished to reduce all of ecosystem effects to species coexistence is that they assumed that our attempt to measure ``higher-order effects'' in mixtures referred exclusively to species interactions. This is indicated by the fact that LH just assumed that the matrix describing species' effects on each other's properties in pairwise mixtures was a Lotka-Volterra model, despite us never having made any reference to ``Lotka-Volterra'' or to any other type of community assembly model. That is akin to seeing a Taylor expansion and assuming that it necessarily entails local stability analysis. Taylor expansion, like matrix inversion, can be used for other purposes: neither technique automatically implies an analysis of the asymptotic behavior of a dynamical model.

Our baseline is grounded in the phenomenological description of how species' \emph{properties} shift in mixtures and is thus independent of any specific type of structuring mechanism or population dynamics. Hence, it could apply to any type of transient or stable community assembled. Furthermore, since our approach tracks the effects on each species' property in a community, \emph{the higher-order effects} it is meant to measure include ecosystem effects, not just the effects of species interactions on population growth observed in Lotka-Volterra models.

Yet LH went so far as to claim that our BIODEPTH numerical simulation results \emph{had} to be wrong because there should be zero residual biodiversity effect produced by a Lotka-Volterra model once the pairwise effects are accounted for unless the model also includes unspecified nonlinearities. However, LH’s claim that our results must be incorrect is simply not true because nowhere in our paper did we state or imply that a Lotka-Volterra model was used to measure ecosystem functioning in communities of three or more species. As we explained in the main text and Appendix S2 of our paper, our BIODEPTH simulation results were obtained by computing the pairwise effects from the relevant 2-species mixture data and then subtracting the sum of these pairwise effects from the degree of ecosystem functioning observed in mixtures of three or more species. Hence, contrary to what was suggested by LH, we never fit any kind of Lotka-Volterra model to mixtures of three or more species; doing so would be tantamount to naively assuming that natural communities operate like Generalized Lotka-Volterra models. Instead, our measure of ecosystem functioning represents higher order effects that are not a direct consequence or corollary of coexistence as adduced from the simplest pairwise mixtures. Our measure can thus include, among other things, biodiversity effects arising from nonlinear interspecific interactions, the effects of non-transitive competition in communities of three or more species, as well as ecosystem level feedbacks.

\subsection*{Null expectations based on principles of coexistence}
Part of LH's misreading stems from them confusing the \emph{abstract principles underlying coexistence theory} that we incorporated into our baseline expectations as to how species pack together, with all of the \emph{particular} factors and mechanisms that actually play out in the real world to affect coexistence. The question then is, what should be the \emph{default expectation} for how species pack together, assuming there is no higher-order (complicating) effects? LH assume that a neutral null community based on perfect redundancy is the appropriate baseline. LH incorrectly claim that their null model is not neutral because they assume that neutrality requires that all species have identical parameter values. However, LH's definition is just an extreme and limiting case of neutrality whereby all individuals in a community are identical. Neutrality has \emph{always} been more broadly defined as interspecific fitness equivalence, which can emerge for species whose demographic parameters differ but trade-off so as to yield the same fitness for all individuals in the community (see \citeauthor{hubbell_01} \citeyear{hubbell_01} page 322; \citeauthor{chave_04} \citeyear{chave_04}; \citeauthor{purves.turnbull_10} \citeyear{purves.turnbull_10}; \citeauthor{munoz_huneman_16} \citeyear{munoz_huneman_16}).

In any case, this disagreement over how the LH null model or baseline should be labeled is largely academic since it does not change what it actually represents: the assembly of a community consisting of a set of species with equivalent (overall) fitness. This means that the LH partitioning method and its baseline always accounted for the effects of neutral coexistence on ecosystem functioning. Hence, the criticisms leveled at us in the \citet{loreau.hector_19} Comment for incorporating the effects of coexistence also apply to the original LH partitioning method! The question then is not \emph{whether} but \emph{what kind} of coexistence should be integrated into the baseline of the LH partitioning method.

Under the LH neutral null model, as species are added, they all share the habitat space equally and thus, \emph{on average}, no augmentation of the total property occurs. We consider this an artificially low baseline because nearly one century of theory suggests that the default expectation should include the possibility of some niche non-overlap. The many ways in which this general abstract coexistence principle actually plays out or is complicated by the nearly infinite set of factors or mechanisms at work in particular communities is not meant to be discounted by our baseline as trivial or uninteresting -- this would defeat the point.

\begin{figure*}[!htb]\centering
\includegraphics[width= 0.72 \linewidth]{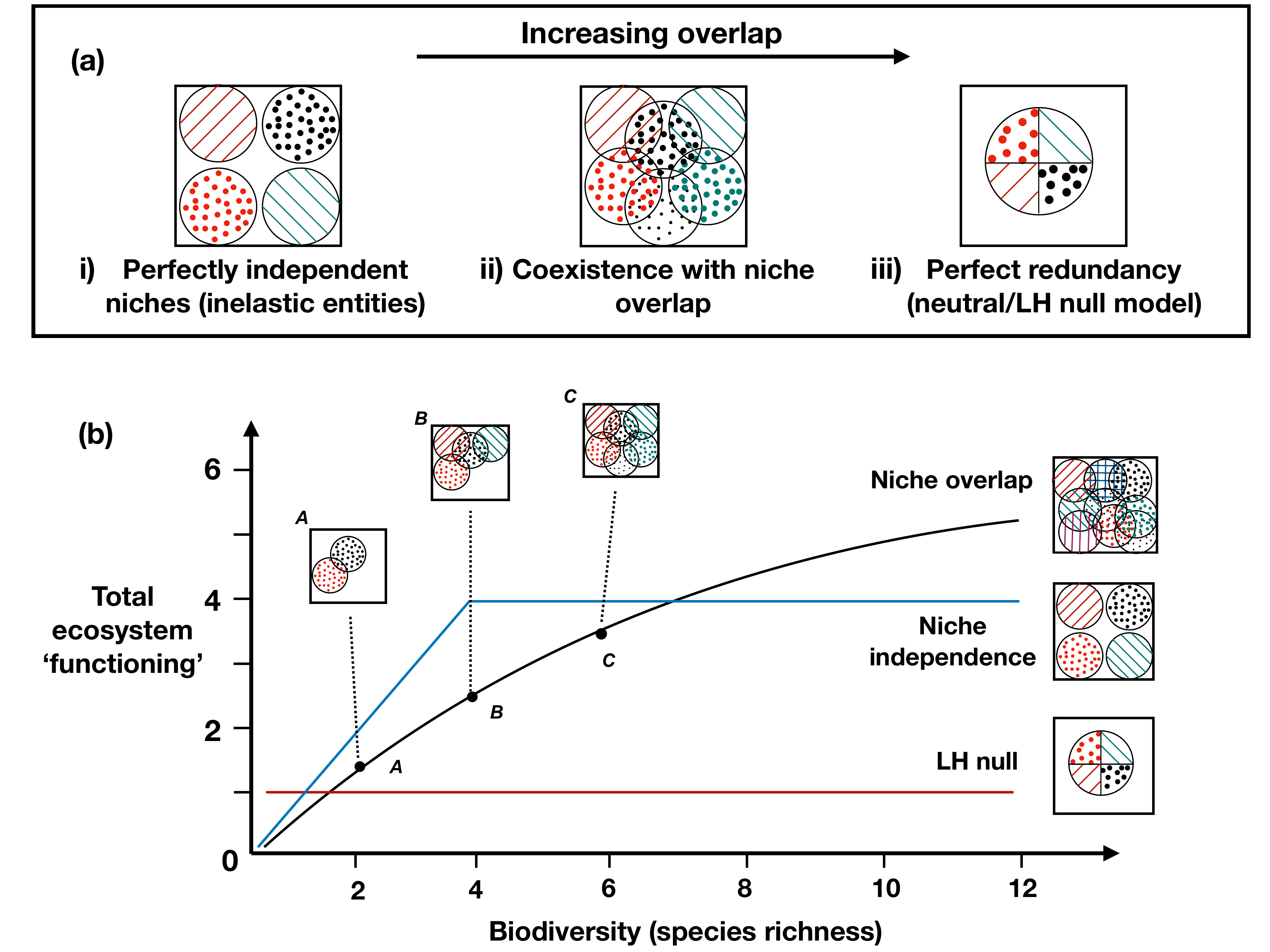}
\caption{\textbf{Range of expectations for species niche packing}. Species niches shown as bodies occupying, to different degrees, the same `space'. Coexistence theory leads to an expectation of some degree of niche non-overlap when species are packed together. For simplicity, all species niches are considered identical in their ecosystem effect (1 species = 1 `ecosystem unit'). (a) Range of default expectations with increasing niche overlap. The LH null model is an extreme and unlikely expectation based on neutrality and perfect redundancy. (b) Ecosystem functioning as biodiversity increases. In the absence of higher-order effects, the degree of niche overlap (how species pack into an abstract volume), will determine how ecosystem properties sum to the total observed. Here the LH null model sets an artificially low threshold for measuring effects based on neutrality/redundancy. However, our metric's baseline for measuring biodiversity effects in high diversity communities is adaptive and can help account for this range of niche overlap, because it establishes a baseline based on empirical observations in simple (2-species) communities. This means that the LH neutral null community arises as a special case of our approach if the experimental evidence from pairwise mixtures warrants it.}
\label{fig1}
\end{figure*}

The problem is how to account for the observed increase in some aggregate property that is merely the corollary of species co-occurring at a given scale whether stably or transiently. For example, increasing the number of inelastic particles or bodies in a container, such as marbles in a jar, will lead to a greater volume being displaced. Yet the question should not be whether the observed total displacement is real or not, but rather, whether such an unsurprising observation constitutes a suitable focus for scientific study, or whether it should be discounted as trivial. A more interesting focus of investigation would be the departures from this expectation, as when the bodies in question could react with each other, transform their immediate environment, or in some other way interact to modify the aggregate property of interest beyond that expected from simply summing up the individual properties associated with individual particles.

Unlike inelastic bodies that cannot occupy the same physical space, species niches are entities that can overlap, at least partially, to occupy the same abstract habitat ‘space' (Fig. \ref{fig1}). But what a century of theory has led us to expect is that, all things being equal, coexisting species are likely to be differentiated with at least some degree of niche non-overlap, and thus, knowing nothing else, our default expectation should be that some aggregate property of a community is likely to correlate (even if only imperfectly) with the number of species that can be packed into a given space. Take a simple thought experiment where we assume that the aggregate ecosystem property in a community is not affected by higher order effects or interactions between species and is thus simply the sum of individual species contributions. Then, as the number of species increases, the change in the total ecosystem property observed will simply be due to the effects arising from species packing (Fig. \ref{fig1}). The issue then becomes how the entities in question (in this case, \emph{species niches}) are expected to pack together. Knowing nothing else, for species niches we would expect some degree of non-overlap, and in the extreme case of independence, completely non-overlapping `inelastic' niches (Fig. \ref{fig1}a). Partial and non-overlapping scenarios for species packing represent the range of cases expected from the simplest and most \emph{idealized} principles of coexistence theory for a century now.

\begin{figure*}[!htb]\centering
\includegraphics[width = 1 \linewidth]{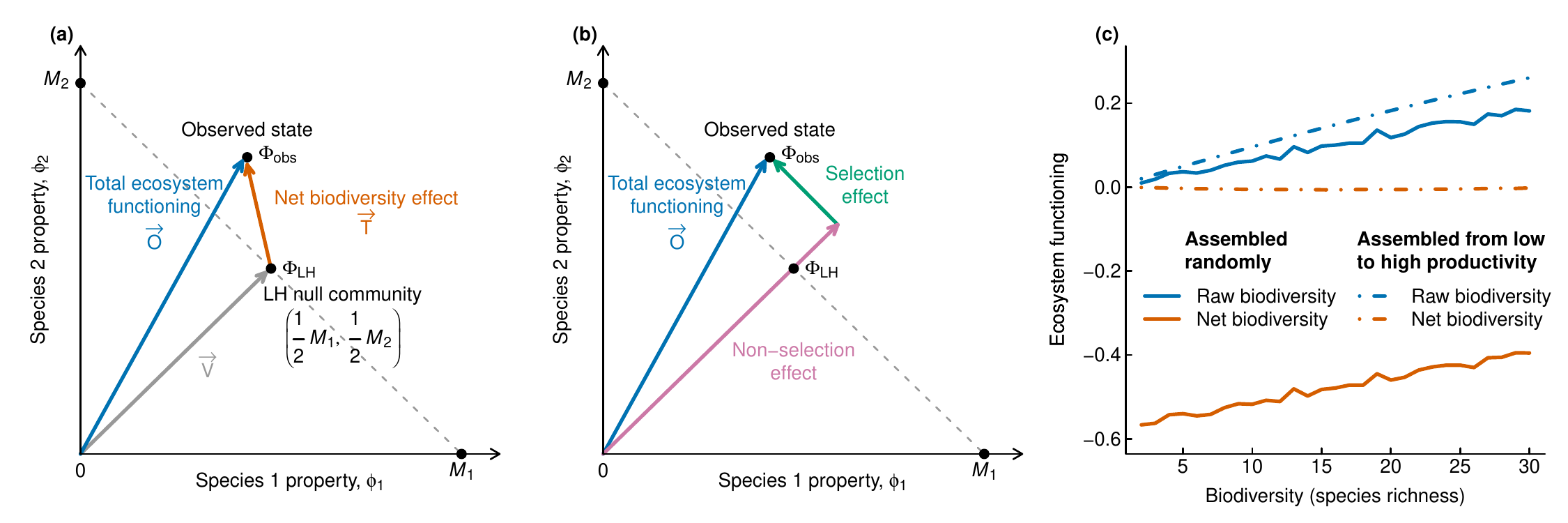}
\caption{\textbf{The net biodiversity effect}. (a) Vector \textbf{T} representing the net biodiversity effect shown as a (non-parallel) portion of the total observed ecosystem property, O. Vector \textbf{V} gives the portion of \textbf{O} deleted when calculating the net biodiversity effect. (b) The selection effect, $n\cdot\mathrm{Cov}[p',M]$, determined by partitioning the total observed ecosystem functioning: $\phi_{\mathrm{obs}} = n\cdot\mathrm{Cov}[p',M] + n\cdot\overline{p'}\;\overline{M}$, where $p' = \frac{O}{M}$
represents the ratio of ecosystem functioning observed in mixtures $O$ to that observed in monocultures $M$. (c) Comparison of the raw and net biodiversity effects for metacommunity simulations with species addition being either fully randomized (solid lines) or based on increasing productivity (dashed lines).}
\label{fig2}
\end{figure*}

In contrast, a baseline where all species are indistinguishable and overlap perfectly, as in the LH neutral model, represents an extreme and highly unlikely coexistence scenario (a `strawman scenario'), and as a result, one that would provide a misleadingly low threshold for measuring biodiversity effects (Fig. \ref{fig1}b). If one wishes, one could certainly use the low baseline provided by the LH null hypothesis; our point was not to question the `reality' of the ecosystem change measured when using it, just its value for scientific research. On the other hand, scientific considerations aside, if one's desire is simply to maximize the chance of measuring a positive BEF relationship in an experiment (say, in order to advocate for the virtues of higher biodiversity), then the LH null model is certainly the baseline to use. The fact that the real world is more complicated than our little \emph{gedankenexperiment} would suggest is precisely the point. Of course we are aware that increased species packing can lead to negative feedbacks and suppression of the total ecosystem property, or alternatively, to positive feedbacks or other positive higher order effects. But it is these departures from expectations based on coexistence theory in its most idealized and abstract form that offer the greatest opportunities to explore how the natural world is structured. In particular, they allow us to disentangle feedbacks and synergistic effects from the effects associated with species packing -- and thus offer the real possibility of exploring the evolution and emergence of ecological complexity.

It appears that LH's caricature of our position, and their inability to appreciate the rationale for our baseline, ultimately stems from a misunderstanding of how our null expectation is motivated by the basic principles of ecological theory. LH failed to recognize that by using our baseline, we deleted the pairwise effects from higher diversity communities as a way to control for the \emph{trivial additive effects} arising from species addition, and that this was not some incomplete or partial way of controlling for \emph{all possible coexistence effects}. These additive effects were considered trivial precisely because they are part of a range of effects that would be expected to arise normally from simply assembling or packing species together in the absence of any complicating factors or mechanisms.

\subsection*{The net biodiversity effect}
What is worse than LH not understanding the reasons why we deleted a fraction of the total ecosystem effect via our baseline is the fact that they do not seem to realize that they too deleted a significant portion of the total ecosystem effect in their null model. Yet, LH state that ``BEF research seeks to identify and understand \emph{all} the ways in which biodiversity affects ecosystem functioning. [...] In this context, the monoculture is the natural baseline since a single species or type corresponds to the presence of life but the absence of biodiversity'' (emphasis added). Contrary to what LH claim, the midpoint (or average) of the monocultures does not merely represent the ``presence of life'' but the assembly of a (theoretical) neutral community that serves as a baseline for computing and partitioning the net biodiversity effect. Just because it is calculated using nothing more than the monoculture yields does not mean that the LH baseline is not equivalent to assembling a neutral community. Hence, there is nothing sacrosanct or special about the midpoint of the monocultures. Calculating the net biodiversity effect relative to this neutral baseline deletes a significant portion of the total ecosystem effect that was implicitly deemed `trivial', whether LH realized it or not (Fig. \ref{fig2}a).

What is perhaps an even more awkward issue for the LH method is that its baseline and the net biodiversity effect itself may have been unnecessary all along. If LH were truly concerned with simply measuring the full effects of biodiversity, then why not use the raw biodiversity effect (i.e., total ecosystem functioning observed in mixtures)? If the purpose of developing the net biodiversity effect was to determine whether increases in ecosystem functioning with diversity were due to sampling or selection effects, then (putting aside nonlinearity) under the standard BEF experimental set-up, LH could have simply partitioned the raw biodiversity effect to obtain the selection effect directly (Fig. \ref{fig2}b). On the other hand, if the purpose of the net biodiversity effect was to determine how biodiversity influences ecosystem functioning by regressing the net biodiversity effect against species richness, then the net biodiversity effect may give a misleading indication of the sign and the magnitude of the biodiversity-ecosystem relationship (e.g., negative or zero net biodiversity effect when the raw biodiversity effect is positive and increases with species richness in a metacommunity assembled via a competition-colonization tradeoff, Fig. \ref{fig2}c). Hence, compared to the raw biodiversity effect, the net biodiversity effect is at best redundant and at worst misleading.

Unlike LH, we are aware that ours is a baseline associated with a null model of community assembly. Consequently, we deliberately chose to delete a portion of the total ecosystem effect in order to isolate what we argue are more interesting effects of biodiversity. LH however seem to believe that their null baseline is a natural measure of the ``presence of life'', and thus devoid of any assumptions regarding community assembly. However, both baselines represent implicitly defined metrics. That is, LH's measure of net biodiversity effect is not a measure of a basic `observable' or a `primitive' quantity like the aggregate ecosystem property, but is defined relative to a set of assumptions, or a theory as to what constitutes a meaningful measure. The problem is that LH simply were not aware of their assumptions -- they took it at as self-evident that one should simply take the monoculture average as the baseline. This is hinted at by the fact that they make no justification for monoculture averages in their original papers, or in their Comment, aside from references to its agricultural origins.

\section*{Claim 2: The effects of nonlinearity}
Even if one remains unconvinced by the conceptual issues described above, the impact of nonlinearity on the foundational analytical tools of BEF, namely LH's net biodiversity effect and additive partitioning, is undeniable. We showed that nonlinearity can artificially inflate the net biodiversity effect and lead to spurious estimates of complementarity and selection effects. Although the LH partitioning remains arithmetically sound (the net biodiversity effect always equals the sum of the selection and the complementarity effects), the individual effect sizes it reports are inaccurate under nonlinearity, thus making the scheme useless for inferential purposes (its \emph{raison d'\^{e}tre}). Despite the critical nature of this issue, LH casually dismissed our mathematical critique as ``superfluous'' and ``less fundamental''. Below we respond to the three offhand points LH made to dismiss our critique.

\subsection*{The LH partitioning is not meant to infer mechanisms}
We were taken aback by LH's surprising assertion that their partitioning was never meant to make inferences. This argument is belied by nearly two decades of definitions, debates and discussions \citep[e.g.,][]{petchey_03} as to the categories of mechanisms that likely contribute to selection effects (SE) and complementarity effects (CE). True, the LH partitioning does not infer specific biological mechanisms or causal effects, and we never claimed (nor believed) that it did. But it is used to infer \emph{categories or groups} of mechanisms that contribute to a given effect, such that a positive CE will mean that ``niche differentiation (partitioning of either resources or natural enemies), positive interactions, or some combination thereof, are strong enough to outweigh interference competition or other negative species interactions that might decrease relative yields in mixture'' \citep{loreau.sapijanskas.ea_12}. This is similar to the definition used in \citet{pillai.gouhier_19}.

The disingenuousness of LH's argument is made clear by the fact that if CE and SE were not meant to correspond to groups of mechanisms or processes, then why label them ``complementarity effects'' and ``selection effects'' instead of ``Effect A'' and ``Effect B''? If we take LH at their word, then we are left with yet another circularity: complementarity effects measure complementarity. What is complementarity? Whatever the complementarity effect measures!

\subsection*{The papers LH cited about nonlinearity}
The closest LH came to addressing our claim that the net biodiversity effect and LH partitioning are mathematically flawed was to cite a series of papers whose only apparent connection to ours is that they all contain the word ``nonlinear'' in them. Unlike these papers, our argument was not about the nonlinear effects arising in mixtures, but about the way nonlinear responses in \emph{monocultures} were confounded with the biodiversity effects measured in mixtures, and how this rendered the LH partitioning meaningless. None of the papers LH cited have any relevance to this argument.

As we mentioned explicitly in our paper (both in the main text and in Appendix S5), nonlinear effects of abundance on ecosystem properties \emph{in mixtures} was never an issue. We stated that the LH method can hold perfectly well even when
\begin{quote}``the \emph{aggregate community-level properties} involve interactive [i.e., nonlinear] effects (Appendix S5). Effects arising from species compositional changes and species interactions (including nonlinear interaction effects) are community-level effects that can be reasonably attributed to diversity. However, BEF studies essentially confound the effects species composition and species interactions have on aggregate ecosystem properties [in species mixtures] with the effects of nonlinear ecosystem-abundance relationships in individual species monocultures.'' \citep[][emphasis added]{pillai.gouhier_19}\end{quote} In short, Loreau and Hector confused the nonlinearity issue by both completely ignoring the detailed mathematical arguments we provided, and by conflating the nonlinearity found in monocultures with that arising in mixtures.

For argument's sake, let us ignore for now this conflating of two distinct types of nonlinearity and pretend that the arguments used by LH to defend their partitioning method are valid.  Consider, then, the claim made in  \citet{loreau.hector_19}'s Comment that our arguments are irrelevant because recent ``developments in BEF research precisely address these more general nonlinearity issues'', and that ``[all] of these new approaches confirmed the significant effects of biodiversity on ecosystem functioning that were found in previous analyses''. What exactly does it mean to say that these new approaches ``confirmed the significant effects of biodiversity''? Again, ignoring the completely different use of the term ``nonlinear'' in this passage from that of our critique, what are Loreau and Hector trying to claim -- that it does not matter that their partitioning does not work accurately because other studies have confirmed that there is some vaguely positive biodiversity-ecosystem functioning relationship? And what does it mean when they go on to state that the LH partition still holds ``qualitatively''? If accurately partitioning some effect size does not matter, then what is the point of having quantitative measures like SE and CE in the first place? Taking Loreau and Hector at their word then, we must assume that they are admitting that all the quantitative measures reported by studies based on the LH partitioning method over the last two decades must now be corrected -- including the ``consensus statement'' by BEF researchers claiming that on average 50\% of biodiversity's effects are due to complementarity and 50\% are due to selection \citep{cardinale.duffy.ea_12}. Ultimately, even if we overlook the fact that the papers cited by LH do not address the nonlinearity issues we uncovered, their attempt to circumvent the issue by vaguely asserting that their partitioning results hold \emph{qualitatively} fails on its own terms.

\subsection*{Linear assumption as a first approximation}
For LH, nonlinearity is not a problem because ``linear models are used as first-order approximations of a more complex, nonlinear reality'', thus, by analogy, dismissing the entire suite of issues raised by nonlinearity as if they were a minor distorting effect in an idealized model, like friction. However, the problem of nonlinearity is one of attribution not estimation; it arises from the inability to (uniquely) map shifts in community space to changes in ecosystem space. It cannot be solved by the familiar ``linearization'' approach that uses a linear function in some arbitrarily small neighborhood around a point to estimate the response of a function that is already approximately linear across the entire relevant domain.

\begin{figure*}[!htb]\centering
\includegraphics[width= 0.72 \linewidth]{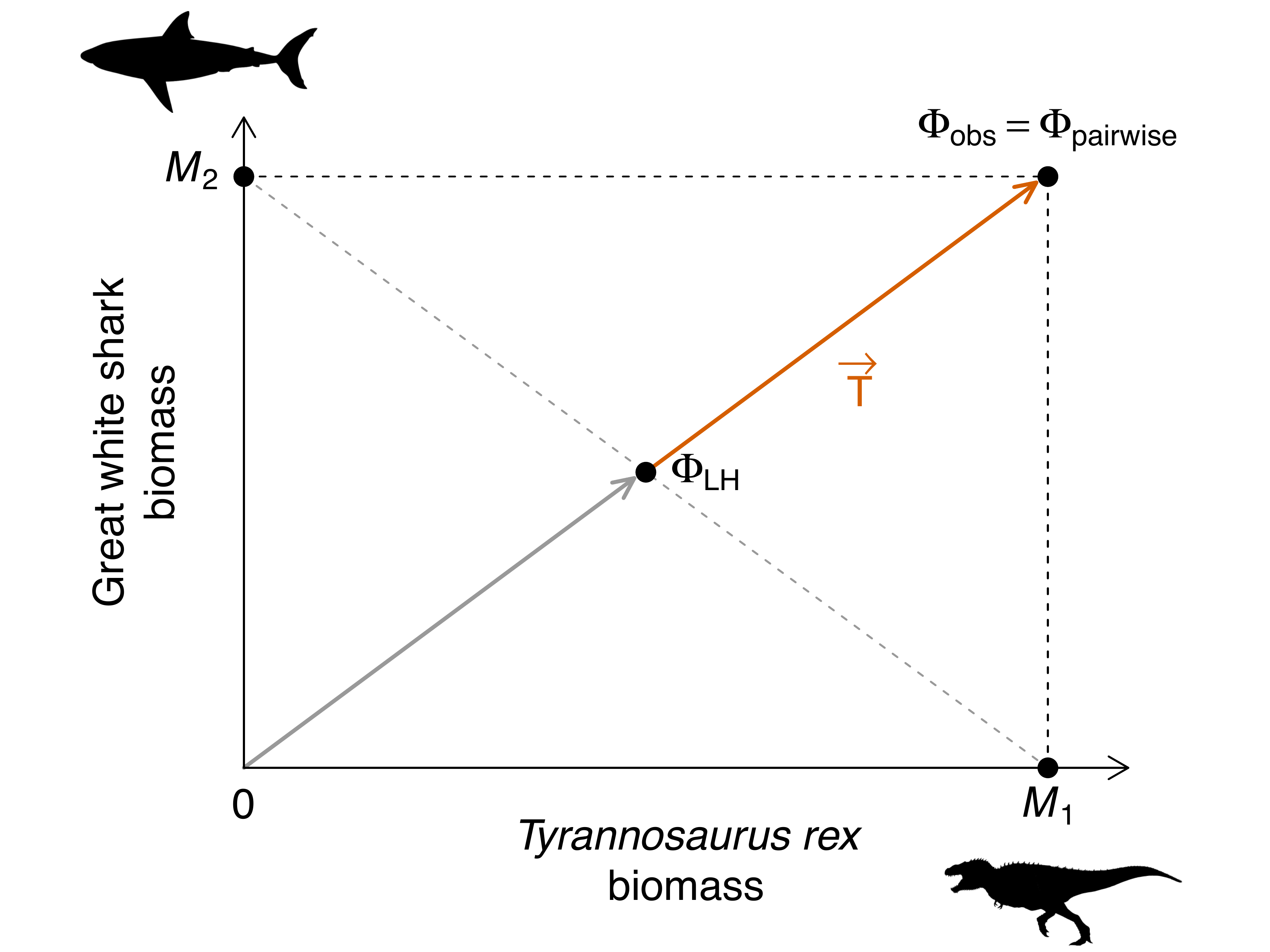}
\caption{\textbf{Quantifying effects of biodiversity on ecosystem functioning across times and places}. `Biodiversity effects' measured across great spatial and temporal scales (spanning K-T extinction) for a guild comprised of \emph{Tyrannosaurus rex} and Great white sharks. Biodiversity's spooky action at a distance is measured by LH's net biodiversity effect, which is given by the vector \textbf{T} (in red) between the LH null, $\mathbf{\Phi_\mathrm{LH}}$, and observed biomass, $\mathbf{\Phi_\text{obs}}$. Our baseline, $\mathbf{\Phi_\text{pairwise}}$, prevents the emergence of this paradox.}
\label{fig3}
\end{figure*}

\section*{Synthesis}
Although our two claims regarding the problems in BEF research are independent, they do converge in ways that highlight the foundational problems in the overall BEF approach. As far as we can see, the only possible way to get around the issues we raised in our original critique is for the BEF community to simply \emph{define away} the problem, as LH to some degree already attempted above with SE and CE. That is, biodiversity effects can simply be defined \emph{a priori} to be \emph{any} changes measured relative to the neutral null baseline community (as given by the net biodiversity effect), thus allowing us to put aside any epistemological issues related to the appropriateness of using such a (low) baseline, or of including nonlinear monoculture effects in our measure of biodiversity effects. However, this will lead to clear paradoxes.

For example, in the case of nonlinearity, consider the situation discussed in Fig. 5 of \citet{pillai.gouhier_19}, where two species (say, plants) in a mixture exist at the midpoint of their carrying capacities. A `biodiversity effect' will be measured if each species exhibits a saturating relationship between abundance and ecosystem functioning in monoculture, even in the absence of any species interactions. Despite the measured effects being purely the result of single-species responses, by using the net biodiversity effect, the sum of these single species responses will now be categorized as a \emph{biodiversity effect}. The problem is that this `biodiversity effect' will exist even if we move the two plants to two separate greenhouses, or even to two greenhouses on different planets – say, Earth and Mars. This `biodiversity effect' will always be observed as it is simply the sum of the individual responses, regardless of the degree of separation. The only reasonable explanation, then, is that some form of ecological `quantum entanglement' allows biodiversity effects to travel \emph{instantaneously} across vast distances: biodiversity's own ``spooky action at a distance''!

A similar spooky action is implied by the use of the LH null model. A positive `biodiversity effect' can easily be measured in an experiment where two species occupying distinct (inelastic) niches are grown in a large mixture (e.g., elephants and blue whales). But the significance of this `biodiversity effect' only holds if we ignore how this effect is an artefact of measuring relative to the LH null community. What is more, biodiversity's spooky action can operate in an even more striking way when we measure biodiversity's effects across space and time. Consider two independent species temporally and spatially partitioning a large habitat, such as \emph{Tyrannosaurus rex} and Great white sharks. Each species in our experiment will grow to the biomasses expected when grown in monocultures, allowing us to see how biodiversity's spooky action can also operate across vast spatial \emph{and} temporal scales (Fig. \ref{fig3}). We trust that most ecologists would consider this paradoxical ``ghost of BEF past'' effect due to spooky action at a distance problematic.

What is important to recall here is that our suggested metric has a \emph{baseline that is adaptive}, and as such, is expected to change for higher diversity mixtures depending on how species have been observed to behave in the simplest pairwise mixtures. In doing so, our baseline will prevent the emergence of the paradoxes and absurdities described above. Additionally, the neutral community model of the LH baseline becomes a special case of our approach whenever the empirical evidence suggests neutral community assumptions are justified for the species involved. Given that our modest proposal for a new baseline is equivalent to that of LH's when 2-species mixtures are neutral, it is difficult to understand how \citet{loreau.hector_19} can suggest that ``what [we] have in mind is an altogether different research program, which indeed requires a different null hypothesis and a different methodology'' and that ``[w]hat [we] do not appear to realize, however, is that this new research program is, first, completely at odds with the BEF research program that [we] criticize and, second, logically inconsistent''. In reality, our baseline
is entirely consistent with the goals of the BEF research program. It merely incorporates (rather than ignores) the basic principles of community ecology
developed over a century ago in order to help identify novel aspects of the biodiversity-ecosystem functioning relationship.

\section*{Conclusion}
We had hoped that our paper would stimulate debates about the difficulties associated with measuring the effects of biodiversity on ecosystem functioning. Unfortunately, our critique based on numerical simulations and mathematical proofs has been met with little more than incoherent argumentation and an irrelevant recounting of the many accomplishments of the BEF research program. Addressing the foundational issues that we identified is critical in order to avoid repeating the mistakes of the past, particularly as BEF research expands to investigate ecosystem multifunctionality across scales. Otherwise, BEF research will continue to bury the interesting effects of biodiversity on ecosystem functioning under layers of triviality while tacitly condoning the absurd phenomena and paradoxes entailed by its assumptions.

%------------------------------------------------
\phantomsection
\section*{Acknowledgments}

We acknowledge support from the National Science Foundation (OCE-1458150, OCE-1635989, CCF-1442728).

%----------------------------------------------------------------------------------------
%	REFERENCE LIST
%----------------------------------------------------------------------------------------
\phantomsection
\bibliographystyle{ecology}
\bibliography{pillai_gouhier_LH_response}

%----------------------------------------------------------------------------------------

\end{document}